\documentclass[aps,prl,twocolumn,longbibliography,showpacs,amsmath,amssymb,amsfonts,nofootinbib]{revtex4-2}
\usepackage{graphicx}
\usepackage{dcolumn}
\usepackage{bm}
\usepackage[usenames,dvipsnames]{color}
\usepackage{multirow}
\usepackage{soul}
\usepackage{natbib}
\usepackage{amsmath, amssymb}
\usepackage{bbold}
\usepackage{physics}
\usepackage{cancel}
\usepackage[hypertexnames=false]{hyperref}

\definecolor{magenta}{rgb}{0.8,0.2,0.8}

\begin{document}

\title{Kardar-Parisi-Zhang universality class in the synchronization of oscillator lattices with time-dependent noise
}

\author{Ricardo Guti\'errez }
\author{Rodolfo Cuerno}
\affiliation{Grupo Interdisciplinar de Sistemas Complejos (GISC), Departamento de Matemáticas, Universidad Carlos III de Madrid, 28911 Legan{\'e}s, Madrid, Spain}

\begin{abstract}
Systems of oscillators subject to time-dependent noise typically achieve synchronization for long times when their mutual coupling is sufficiently strong. The dynamical process whereby synchronization is reached can be thought of as a growth process in which an interface formed by the local phase field gradually roughens and eventually saturates. Such a process is here shown to display the generic scale invariance of the one-dimensional Kardar-Parisi-Zhang universality class, including a Tracy-Widom probability distribution for phase fluctuations around their mean. This is revealed by numerical explorations of a variety of oscillator systems: rings of generic phase oscillators and rings of paradigmatic limit-cycle oscillators, like Stuart-Landau and van der Pol. It also agrees with analytical expectations 
derived under conditions of strong mutual coupling. The nonequilibrium critical behavior that we find is robust and transcends the details of the oscillators considered. Hence, it may well be accessible to experimental ensembles of oscillators in the presence of e.g.\ thermal noise.

\end{abstract}


\maketitle

\noindent {\it Introduction---} Critical dynamics, i.e., the time evolution of a spatially-extended system at a critical point, is paramount throughout physics \cite{tauber14}, biology \cite{Munoz2018}, and society \cite{Jusup2022}. Recent examples range from active matter \cite{Nardini2017,Caballero2020} to driven-dissipative Bose condensates \cite{fontaine,Sieberer2023}, quantum magnetism \cite{wei22,Rosenberg24}, or Anderson localization in disordered conductors \cite{Mu2024}. One key player in this context is the stochastic Kardar-Parisi-Zhang (KPZ) equation with time-dependent noise \cite{kardar}, 
which actually displays generic scale invariance (GSI) regardless of parameter values \cite{tauber14,grinstein95}. Having been originally put forward as a versatile model \cite{barabasi,halpinhealy,krug97} for the growth of a rough interface or, via suitable mappings, directed polymers in random media, or randomly stirred fluids, this system represents a widely-encompassing universality class \cite{takeuchi} that transcends e.g.\ the classic models A or B of critical dynamics \cite{tauber14} far from equilibrium. Beyond some of the previous examples \cite{fontaine,Sieberer2023,wei22,Mu2024}, the KPZ class further includes instances of e.g.\ turbulent liquid crystals \cite{Takeuchi2011}, thin film growth \cite{Almeida14}, random geometry \cite{santalla15}, active fluids \cite{altman15}, or quantum entanglement \cite{nahum17}. A remarkable trait of the one-dimensional (1D) KPZ universality class is that the fluctuation statistics is described by the Tracy-Widom (TW) family of probability distribution functions (PDF) \cite{kriecherbauer10,takeuchi}. Indeed, these PDFs have recently been found in systems across different length scales irrespective of their physical nature \cite{makey20}, providing analogs of the Gaussian distribution for strongly correlated variables \cite{Wolchover2018}.


A seemingly different, albeit no less paradigmatic, phenomenon is the synchronization of assemblies of coupled oscillators \cite{pikovsky,osipov,boccaletti_book}. Indeed, this dynamical process bears crucial importance for the collective performance of many systems throughout science and technology, from coupled chaotic lasers, to swarms of fireflies, spiking neurons, or pacemaker cells in the heart \cite{pikovsky,osipov,boccaletti_book,boccaletti,acebron,arenas}. Traditionally, the time evolution of an oscillator array into its synchronized state has been implicitly assumed to be system-specific and nonuniversal. Nevertheless, connections between synchronization and the KPZ equation have been noted along the years \cite{kuramoto_book,pikovsky,Munoz03}. 
They have been underscored more recently due to e.g.\ the relation between synchronization models and the so-called compact KPZ equation (see e.g.\ Ref.\ \cite{Sieberer18} and others therein), relevant to quantum systems like exciton-polariton condensates \cite{Sieberer2023}. However, only in the case of idealized oscillator models has the synchronous dynamics been partially shown to display 1D KPZ universal properties or related features \cite{lauter,moroney,Moroney23}. If the natural frequencies of the oscillators are disordered and quenched, both the scaling ansatz and the scaling exponent values characterizing the critical dynamics of the synchronization process have been shown to actually differ from 1D KPZ, agreeing, rather, with those of the KPZ equation with columnar disorder \cite{szendro,Lepri2022}, for 1D lattices of both, idealized phase oscillators \cite{gutierrez} and of more realistic limit cycle oscillators \cite{gutierrez2}. Nevertheless, fluctuations are TW-distributed in all cases \cite{gutierrez,gutierrez2}. Yet elucidating the full relation of synchronization in the presence of time-dependent noise to 1D KPZ remains crucial, as such noise occurs in many experimental systems where synchronization emerges \cite{acebron,Lindner04}.

In this paper, we show that the space-time process whereby oscillators readjust their frequencies and eventually synchronize displays, for 1D systems with time-dependent noise, the full set of universal properties of the 1D KPZ universality class, i.e., the scaling ansatz, the critical exponent values, and the TW statistics. This is borne out by an extensive numerical analysis of large systems of phase oscillators, and of two paradigmatic limit-cycle oscillators, namely those of Stuart-Landau (SL) and van der Pol (vdP). It is only for odd couplings among the phases (a symmetry unlikely to hold exactly in the phase-reduced dynamics of generic limit-cycle oscillators, yet present in the celebrated Kuramoto model \cite{kuramoto_book,acebron}) that the critical behavior differs, as it belongs to the Edwards-Wilkinson (EW) universality class, i.e., that of the linearized KPZ equation \cite{barabasi,halpinhealy,krug97}. 
The robustness of our results against changes in the nature of the oscillators suggests a potential relevance for experiments, and also that of sychronization to critical dynamics at large.


\noindent {\it Models---} We consider the synchronization of rings of three different self-sustained oscillators in the presence of time-dependent noise. In all cases the dynamics reads
\begin{equation}
\dot{\bf r}_j = f({\bf r}_j) + K \Gamma({\bf r}_{j+1}-{\bf r}_j) + K \Gamma({\bf r}_{j-1} - {\bf r}_j) + {\bf \xi}_j ,
\label{geneq}
\end{equation}
for $j=1,2,\ldots, L$, with periodic boundary conditions (PBC), i.e.\! ${\bf r}_{L+1} \equiv {\bf r}_1$ and ${\bf r}_0 \equiv {\bf r}_L$, where the state vector ${\bf r}_j \in \mathbb{R}^n$ (here, $n= 1$ or $2$) and dot is time derivative. The autonomous evolution of each dynamical unit is given by the vector field $f$ 
(see below), and is always chosen so that it presents a stable limit cycle with period of oscillation $2\pi$ (unit angular frequency). The coupling is diffusive \cite{osipov} and given by an overall coupling strength $K \geq 0$ and a coupling function $\Gamma$, 
whose range is $\mathcal{O}(1)$. Some of the components $\xi_j^{(a)}$, $a =1,2,\ldots,n$, of the noise terms may be set to zero in some particular system. The nonzero noise components are independent and Gaussian, with zero mean and variance $D$, and delta-correlated in time, $\langle \xi_j^{(a)}\!(t)\, \xi_k^{(b)}\!(t')\rangle = D\, \delta_{jk}\, \delta_{a b}\, \delta(t-t')$. Our simulations employ the standard Euler-Maruyama scheme \cite{Toral} with $0.01$ time step, and are shown for $K=1$ and $\sqrt{D} = 0.1$. 

The first dynamical system that we consider is the Kuramoto-Sakaguchi (KS) phase oscillator \cite{sakaguchi86}. The state variable ${\bf r}_j$ (phase) 
is a scalar, henceforth denoted $\phi_j$, 
while $f(\phi_j) = 1$. The coupling function is $\Gamma_\text{KS}(\Delta\phi) = \sin(\Delta \phi+ \delta)$, 
where $\Delta \phi$ is the phase difference between two neighboring oscillators and the parameter $\delta\in \mathbb{R}$. 
There is a single noise component ($n=1$).

Our second dynamical system is the SL oscillator. This limit-cycle oscillator provides the universal amplitude equation 
at a supercritical Hopf bifurcation \cite{strogatzbook}. The amplitude has two components, described through $z_j \in \mathbb{C}$. The autonomous dynamics, $f(z_j) = (\mu + i)\, z_j(t) - |z_j(t)|^2 z_j(t)$, has an attracting, circular limit cycle with radius $\sqrt{\mu}$, for $\mu>0$, with $\mu \in \mathbb{R}$ measuring distance to bifurcation. The coupling function is $\Gamma_\text{SL}(\Delta z) = c\, \Delta z$, where $\Delta z$ is the state difference between two neighboring oscillators and $c = (1+i \gamma)/\sqrt{1+\gamma^2} \in \mathbb{C}$ depends on parameter $\gamma\in \mathbb{R}$. Two additive noise components affect ${\rm Re} \, z(t)$ and ${\rm Im} \, z(t)$.

We finally consider the vdP oscillator, a full-fledged limit-cycle oscillator of experimental relevance, introduced in the context of electronic circuits almost a century ago \cite{strogatzbook}. The autonomous dynamics reads $\ddot{x}_j = (\mu - x_j^2)\, \dot{x}_j - x_j$ which, by defining ${\bf r}_j = (x_j,v_j \equiv\dot{x}_j)$, 
implies $f(x_j, v_j) = (v_j, (\mu - x_j^2)\, v_j - x_j)$. In the absence of coupling and noise, a stable limit-cycle ensues with ${\cal O}(\sqrt{\mu})$ size for $\mu>0$, above a supercritical Hopf bifurcation at $\mu = 0$ \cite{strogatzbook}. The coupling function is $\Gamma_\text{vdP}(\Delta x, \Delta v) = (0,(\gamma \Delta x + \Delta v)/\sqrt{1+\gamma^2})$, where $\Delta x$, $\Delta v$ are the components of the state difference between two neighboring oscillators, and $\gamma$ plays a role analogous to the homonymous SL parameter. 
Noise only enters the equation giving $\dot{v}_j$.
For SL and vdP oscillators, we inspect $\mu = 0.5$ and $1$ in our simulations below, to explore the effect of distance from the bifurcation point.

\begin{figure}[t!]
\includegraphics[scale=0.35]{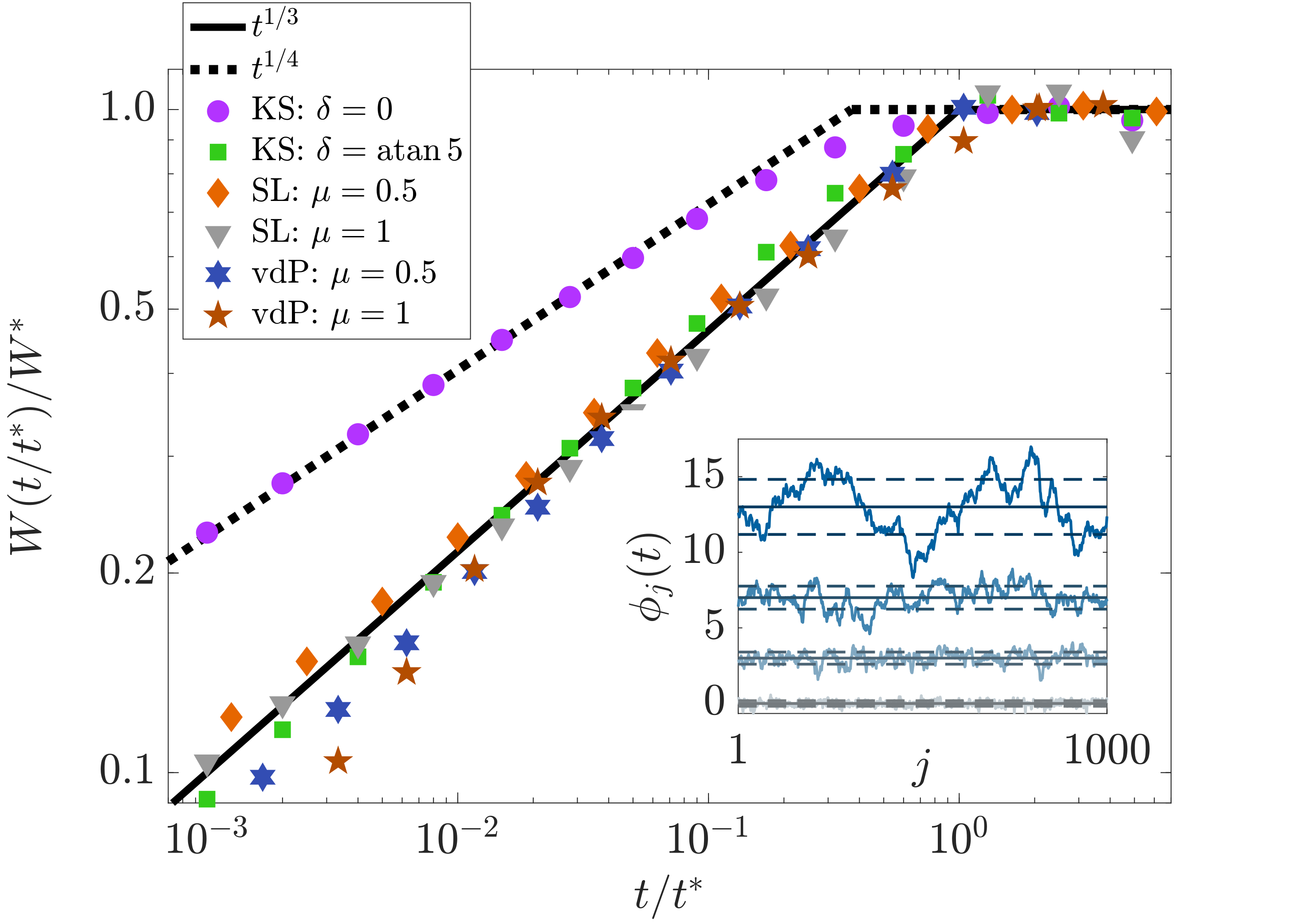}
\vspace{-0.5cm}
\caption{{\sf \bf Roughness of phase profiles in rings of self-sustained oscillators.}
({\it Main panel}\,)  Roughness (\ref{W}) normalized by its saturation value, $W/W^*$, as a function of time rescaled by the saturation time, $t/t^*$, for rings of $L=1000$ oscillators of all three dynamical systems under consideration (see legend). While the KS rings with $\delta = 0$ display growth exponent $\beta = \beta_{\rm EW} = 1/4$, all other rings of oscillators display $\beta = \beta_{\rm KPZ} = 1/3$. 5000 noise realizations have been used in the averaging. ({\it Inset}) Phases $\phi_j(t)$ in a ring of $L=1000$ KS oscillators with $\delta = \atan 5$, forming a phase profile that becomes progressively rougher, at rescaled times $t/t^* = 10^{-3}, 10^{-2}, 10^{-1}$. and $1$ (bottom to top). Solid (dashed) lines show the space average (one standard deviation above and below average, 
as an estimate of the roughness). Data have been shifted vertically for visualization purposes.}
\label{Fig1}
\end{figure}

\noindent {\it Phase profiles---} To connect with KPZ physics, we describe the state of the oscillators along the ring by a set of phases $\left\{\phi_j(t)\right\}_{j=1}^L$, playing an analogous role to that of the local surface heights above a substrate, forming what we term a ``phase interface''; see an example in Fig.~\ref{Fig1} (inset) for the simple KS model. 
The dynamics of limit-cycle oscillators 
can also be described by the evolution of a single phase variable \cite{pikovsky}. Moreover, close enough to the bifurcation point one can perturbatively approximate such systems purely by phase dynamics {\em via} phase-reduction techniques \cite{kuramoto_book,pietras}. 
For SL oscillators, the uncoupled ($K=0$) noiseless dynamics is characterized by the stable limit-cycle solution $z_j(t) = \sqrt{\mu} \exp{i \theta(t)}$, the (unwrapped) geometric phase $\theta_j(t) = t + \theta_j(0)$ \cite{introsl} acting as a valid dynamical phase variable \cite{pikovsky}. Off the limit cycle, 
an asymptotic phase \cite{kuramoto_book} can be explicitly defined 
as $\phi_j(t) = \phi(z_j(t)) = \theta_j(t)$, i.e.\! the dynamic phase is the geometric phase everywhere in phase space (again considering the unwrapping of $2\pi$ oscillations) \cite{nakao}. 
Thus the instantaneous phases $\{\phi_j(t)\}_{j=1}^L$ are readily extracted from the states of the SL oscillators $\{z_j(t)\}_{j=1}^L$, on or off the limit cycle. A similar behavior is expected for vdP oscillators with $\mu\ll 1$. However, as $\mu$ is increased, vdP oscillators behave progressively as relaxation oscillators, moving at different speeds in different regions of non-circular limit cycles \cite{strogatzbook}. 
We thus define the phases $\{\phi_j(t)\}_{j=1}^L$ differently, namely by the crossing times of a Poincar\'e section \cite{pikovsky}; see the Supplemental Material (SM) for details \cite{sm}.

\noindent {\it Scale-invariant roughening---} The deviations of the local phases around their mean are measured by the roughness,
\begin{equation}
W(t) \equiv \langle \overline{[\phi(x,t)-\overline{\phi(x,t)}]^2} \rangle^{1/2},
\label{W}
\end{equation}
with the overbar denoting spatial average 
and the brackets denoting average over different noise realizations \cite{barabasi}. 
As we focus on large system sizes $L\gg 1$, we employ a continuum notation for the phase field, $\phi(x,t)$. 
The roughness saturates if the phase differences remain bounded in time, which only happens when all oscillators reach a common effective frequency. Thus, $W$ reaching a steady state value manifests synchronization (i.e.\ frequency entrainment) \cite{gutierrez, gutierrez2}, which occurs for coupling $K > K_c$ above a synchronization threshold $K_c$. 

Under the standard kinetic-roughening conditions \cite{barabasi,halpinhealy,krug97} which extend equilibrium critical dynamics \cite{tauber14} to a host of other systems, local degrees of freedom are statistically correlated for distances below a time-increasing correlation length $\xi(t) \sim t^{1/z}$, where $z$ is the dynamic exponent. After a saturation time $t^* \sim L^z$, $\xi(t)$ is comparable to $L$, the roughness saturating to a steady-state value $W^* \sim L^\alpha$, where the roughness exponent $\alpha$ characterizes the fractality of the profile \cite{barabasi,Mozo22}. Prior to that, in the growth regime, the roughness increases with time as $W(t)\sim t^{\beta}$, where the growth exponent $\beta=\alpha/z$. 
This behavior is indeed found in our phase profiles, as seen in Fig.~\ref{Fig1} (main panel), which shows the roughness as a function of time rescaled by saturation values, $W(t/t^*)/W^*$, for rings of $L=1000$ oscillators for the KS model with $\delta = 0$ and $\delta = \atan 5$ (see SM for justification of this choice \cite{sm}), and for the SL and the vdP oscillators. 
Here $t^*$ is estimated as the time when $W$ appears to reach the steady state, and $W^*$ as the time average of $W$ restricted to $t\geq t^*$. For the KS model with $\delta = 0$, the growth exponent follows closely that of the EW universality class, $\beta_{\rm EW} = 1/4$, while in any other case it appears to be that of the KPZ universality class, $\beta_{\rm KPZ} = 1/3$, with some deviations for vdP oscillators at small scales. While these results are robust against variations of the coupling strength as long as there is synchronization, for couplings below the synchronization threshold, $K<K_c$, the resulting (nonsynchronous) process appears to behave as random deposition in all cases, as expected for the KS model with $K=0$ (see the SM for details \cite{sm}).

\begin{figure}[t!]
\includegraphics[scale=0.33]{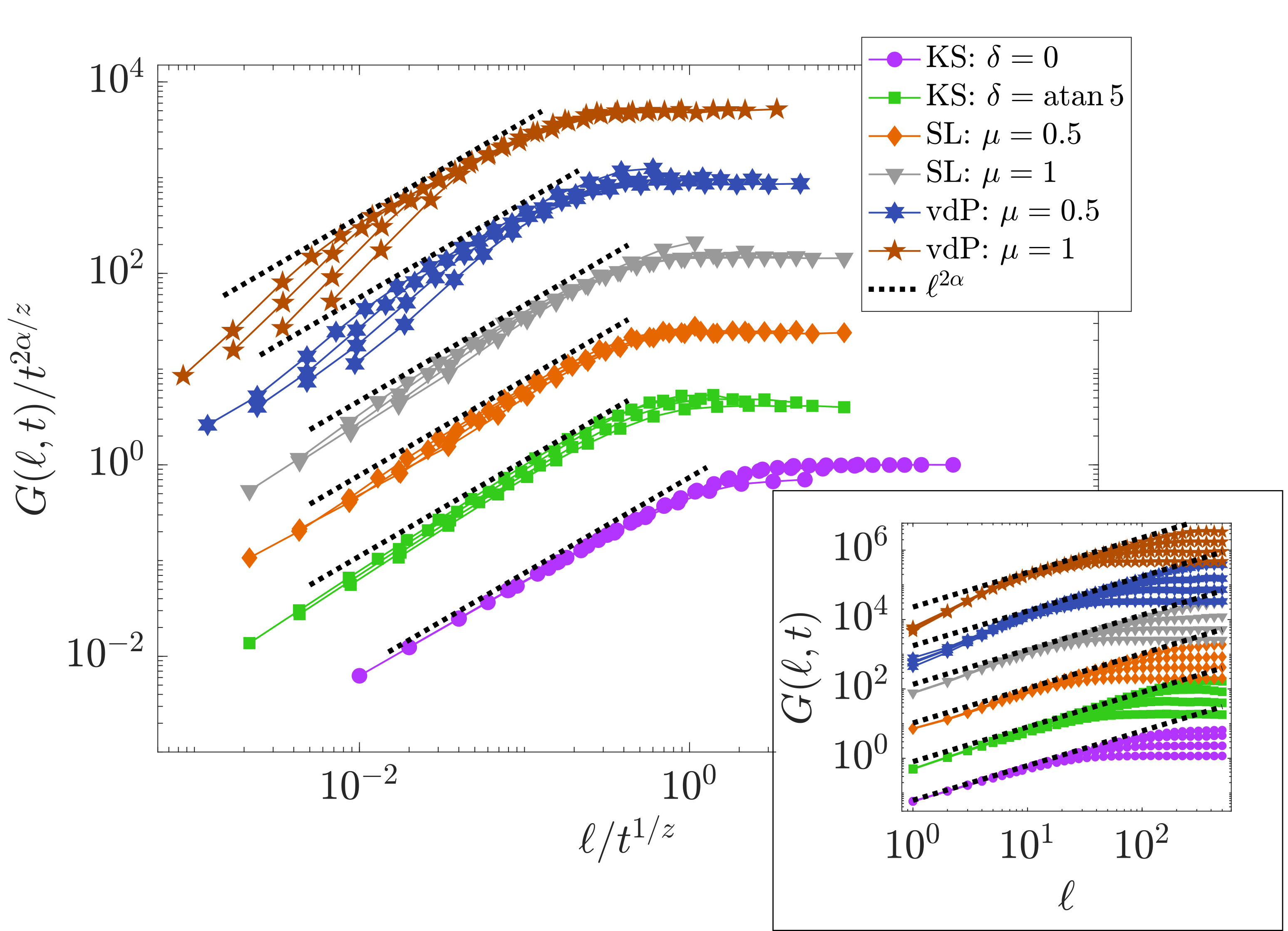}
\caption{{\sf \bf Phase-difference correlation functions of rings of self-sustained oscillators.}
({\it Main panel}\,) Collapse as in Eq.~(\ref{glt}) of the data displayed in inset for the phase-difference correlations $G(\ell,t)$, Eq.~(\ref{grt}), for rings of KS, SL, and vdP oscillators, with the same parameters as in Fig.~\ref{Fig1} (see legend). Exponent values are $\alpha = \alpha_{\rm KPZ} = \alpha_{\rm EW} = 1/2$ (dashed lines represent $\ell^{2\alpha}= \ell$), in all cases, and $z = z_{\rm KPZ} = 3/2$, except for the KS ring with $\delta = 0$, for which $z=z_{\rm EW} =2$.  ({\it Inset}) Phase-difference correlation function $G(\ell,t)$, Eq.~(\ref{grt}), at rescaled times $t/t^* = 1/8^z, 1/4^z, 1/2^z$, and $1$ (bottom to top).  An arbitrary vertical shift has been used for visualization purposes in both panels. Results based on $5000$ realizations.} 
\label{Fig2}
\end{figure}

\noindent {\it Correlations---} Beyond global quantities like $W$, kinetic roughening also reveals itself in local correlation functions like the phase-difference correlation,
\begin{equation}
    G(r,t) \equiv \langle \overline{[\phi( x+ r,t) - \phi(x,t)]^2} \rangle ,
\label{grt}
\end{equation}
which follows the Family-Vicsek (FV) dynamic scaling ansatz \cite{Family1985,barabasi,halpinhealy,krug97} 
\begin{equation}
    G(\ell,t) \sim \left\{
    \begin{array}{ll}
        t^{2 \beta},& \text{if } t^{1/z} \ll \ell, \\
        \ell^{2 \alpha}, & \text{if } \ell  \ll t^{1/z}
    \end{array}
    \right\} = t^{2\beta} g(\ell/\xi(t)) ,
\label{glt}
\end{equation}
where $\ell \equiv |r| <L$, $g(y) \sim y^{2\alpha}$ for $y\ll 1$, and $g(y) \sim {\rm cnst.}$ for $y\gg 1$. 
The scaling form, Eq.\ (\ref{glt}), reflects the same growth and saturation of $W(t)$, restricted to a region of linear size $\ell$ \cite{lopezphysa}. In Fig.~\ref{Fig2} (inset) we show $G(\ell,t)$ for the same systems considered in Fig.~\ref{Fig1}.
The validity of the FV ansatz is assessed in the main panel of Fig.~\ref{Fig2}, where it yields excellent data collapse, again except for vdP oscillators at small scales. Consistent with Fig.\ \ref{Fig1}, the exponent values are 1D KPZ for all cases, $\alpha_{\rm KPZ} = 1/2$ and $z_{\rm KPZ} = 3/2$, except for the KS model with $\delta = 0$, where EW scaling holds, i.e., $\alpha_{\rm EW} = 1/2$ and $z_{\rm EW}=2$. Further evidence on exponent values comes from Fourier-space correlations measured by the structure factor \cite{siegert96,lopezphysa}; see the SM \cite{sm}.

\begin{figure}[t!]
\includegraphics[scale=0.35]{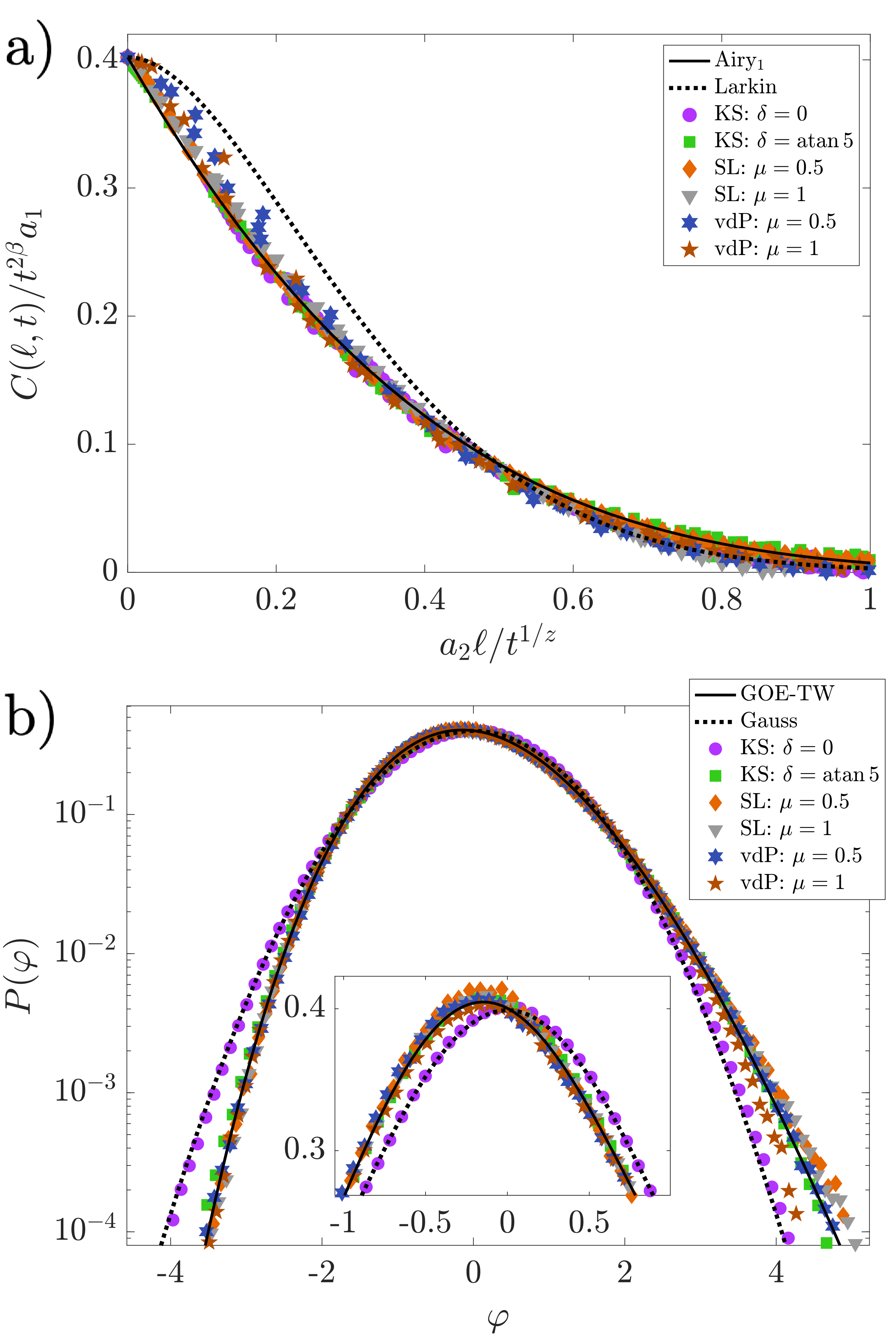}
\caption{{\sf \bf Phase covariance and phase fluctuations of rings of self-sustained oscillators.}
(a) Rescaled phase covariance, Eq.~(\ref{eq:cov}), for rings of $L=10000$  KS, SL, and vdP oscillators with parameters as in Fig.\ \ref{Fig1}. The solid (dotted) line represents the covariance of the Airy$_1$ process (Larkin model). 
(b) Histogram of phase fluctuations, Eq.~(\ref{fluct}), in semi-logarithmic scale for same rings of oscillators (see legend). Inset: Same data in linear scale, centered around the peak of the distribution. The solid (dotted) line represents a GOE-TW (Gaussian) PDF. 
Results based on $1000$ realizations, with time points $t= t_0 + \Delta t$ equispaced along the growth regime, and $t_0$ at its initial stages. Exponent values are as in Fig.\ \ref{Fig2}.}
\label{fig3}
\end{figure}

For 1D KPZ, two-point correlations actually take a universal form, known for the phase covariance \cite{kriecherbauer10,takeuchi}
\begin{equation}
C(r,t) \equiv \langle \overline{\phi(x,t) \phi(x+r,t)} \rangle - \langle \bar{\phi}(t) \rangle^2 ,
    \label{eq:cov}
\end{equation}
which is shown in Fig.~\ref{fig3} (a) for the same systems and parameters. In the growth regime \cite{kriecherbauer10,takeuchi}, $C(\ell,t) = a_1 t^{2\beta} \mathcal{C}(a_2 \ell/t^{1/z})$, where $\mathcal{C}$ is a scaling function, and $a_1$ and $a_2$ are constants \cite{barreales2020}. The covariance of the Airy$_1$ process, expected for 1D KPZ and EW with PBCs \cite{takeuchi,carrasco}, indeed occurs for all our systems, including the KS model with $\delta =0$.
The covariance of the Larkin model \cite{purrello}, also shown, applies to the same oscillator systems, but in the presence of columnar disorder \cite{gutierrez, gutierrez2}.

\noindent {\it Fluctuation statistics---}
A final remarkable trait of 1D KPZ is the universality of the fluctuations \cite{kriecherbauer10,halpinhealy,takeuchi}. These are here studied in terms of rescaled phases,
\begin{equation}
\varphi_j\equiv \frac{\delta \phi_j(t_0+\Delta t) - \delta \phi_j(t_0)}{(\Delta t)^\beta} ,
\label{fluct}
\end{equation}
where $\delta \phi_j(t) = \phi_j(t) - \overline{\phi}(t)$, $t_0$ is a reference time beyond the initial transient dynamics, and $t_0 +\Delta t$ lies within the growth regime. Division by $(\Delta t)^\beta$ removes the systematic time increase of fluctuations so that, remarkably, the PDF of $\varphi_j$ reaches a universal, time-independent form \cite{kriecherbauer10,halpinhealy,takeuchi}. This is Gaussian for EW and the TW PDF associated with the Gaussian Orthogonal Ensemble (GOE-TW) for the KPZ equation with PBCs. Fig.~\ref{fig3} (b) shows the histogram of fluctuations defined as in Eq.~(\ref{fluct}) in semi-logarithmic scale (main panel) and linear scale focusing around the distribution peak (inset). Again our numerics provide strong evidence that rings of KS oscillators with $\delta = 0$ are in the 1D EW universality class, while the rest of systems under considerations are in the 1D KPZ universality class. The only systematic deviations appear in the right tail for vdP oscillators sufficiently far from bifurcation, $\mu = 1$, at probabilities smaller than $10^{-3}$, which might be related to the difficulty inherent in the definition of a dynamical phase for such systems.

\noindent {\it Continuum approximation: phase interface---} We can rationalize the above findings by considering a relatively slow spatial variation of the phases, as for $K \gg K_c$, for a ring of oscillators with lattice constant $a$. A continuum approximation \cite{gutierrez} which neglects terms $o(a^2)$ yields 
\begin{equation}
\partial_t \phi(x,t) =   \nu\, \partial_x^2 \phi(x,t) + \frac{\lambda}{2}\, [\partial_x \phi(x,t)]^2 + \text{noise},
\label{eqkpz}
\end{equation}
where $\nu \equiv a^2\, \Gamma_\phi^{(1)}(0)$, $\lambda/2 \equiv a^2\, \Gamma_\phi^{(2)}(0)$, $\Gamma_\phi(\Delta \phi)$ is the coupling function within phase reduction \cite{pietras,gutierrez2}, and $\Gamma_\phi^{(k)}(0)$ its $k$-th derivative at $\Delta\phi=0$. For time-dependent noise, Eq.\ \eqref{eqkpz} is the 1D KPZ equation, where $\nu$ represents surface tension and $\lambda$ growth along the local normal direction \cite{barabasi,halpinhealy,krug97}. Detailed expressions of $\nu$ and $\lambda$ are provided at the SM \cite{sm} for the oscillator systems we study. In particular, for the KS model we obtain $|\lambda/2\nu| = \tan \delta$ \cite{sm,[{$\nu$ is positive in all cases under consideration; $\lambda$ may be positive or negative, which amounts to a choice of upward or downward-pointing normal as the preferential local growth direction along the phase interface}][{}]test}, which is zero [i.e., Eq.\ \eqref{eqkpz} becomes the EW equation] for $\delta=0$ (Kuramoto model \cite{kuramoto_book,acebron}), in conformity with our numerical results. Concerning both SL and vdP oscillators, we find \cite{sm} $|\lambda/2\nu| = \gamma \not\equiv 0$, suggesting KPZ behavior as we numerically obtain \cite{note2,forrest,notecoeffs}.

\noindent{\it Conclusions---} 
A numerical study of rings of self-sustained oscillators with time-dependent noise reveals that their dynamical evolution into synchronization features GSI with all the traits of the 1D KPZ universality class, including TW statistics. 
A special case is the KS model with $\delta =0$ (the celebrated Kuramoto model), which displays EW behavior. 
These conclusions agree with analytical expectations from a continuum approximation for large coupling $K$. The behavior of the Kuramoto model seems crucially related to the odd symmetry of its coupling function 
\cite{ostborn}, leading to an up-down symmetry violated in the other cases \cite{gutierrez}. The critical dynamics thus elucidated opens up questions \cite{barabasi,tauber14}, to be addressed for synchronization in general, concerning critical dimensions, (ir)relevant interactions, the role of symmetries, etc.  Moreover, since for columnar disorder synchronization generically displays the critical exponents and anomalous scaling ansatz of the columnar KPZ equation (the columnar EW equation for the KS model with $\delta = 0$), with a TW PDF 
and a Larkin covariance 
\cite{gutierrez,gutierrez2}, intriguing questions arise on the interplay between this type of disorder 
and time-dependent noise. 

As synchronization in 1D oscillator systems affected by time-dependent noise appears to display robust critical features generically in the KPZ class, their observation in experiment (on, e.g., electronic circuits, oscillating chemical reactions, or electrochemically active neurons or heart cells
\cite{pikovsky,osipov,boccaletti_book}) remains an alluring possibility.

\begin{acknowledgments}
\noindent {\it Acknowledgements--} This work has been partially supported by Ministerio de Ciencia e Innovaci\'on (Spain), by Agencia Estatal de Investigaci\'on (AEI, Spain, 10.13039/501100011033), and by European Regional Development Fund (ERDF, A Way Of Making Europe) through Grants No.\ PID2021-123969NB-I00 and No.\ PID2021-128970OA-I00.
\end{acknowledgments}

\onecolumngrid
\newpage

\renewcommand\thesection{S\arabic{section}}
\renewcommand\theequation{S\arabic{equation}}
\renewcommand\thefigure{S\arabic{figure}}
\setcounter{equation}{0}
\setcounter{figure}{0}

\begin{center}
{\large{\bf SUPPLEMENTAL MATERIAL:\\
Kardar-Parisi-Zhang universality class in the synchronization of oscillator lattices with time-dependent noise}}
\end{center}

\maketitle

In the first section of this document we give additional information on the extraction of the phases of one of the three systems discussed in the main text; in the second one, we provide details on the continuum approximation and the choice of parameters; in the third one, we show the dependence of the roughenss on the coupling strength, including values below the threshold for synchronization; in the fourth one, we consider the phase structure factor (two-point correlation function in Fourier space) for all systems under study, which complements the real-space correlation results of Fig.~2 in the main text.

\section{Rings of van der Pol oscillators}

\begin{figure}[h!]
\includegraphics[scale=0.40]{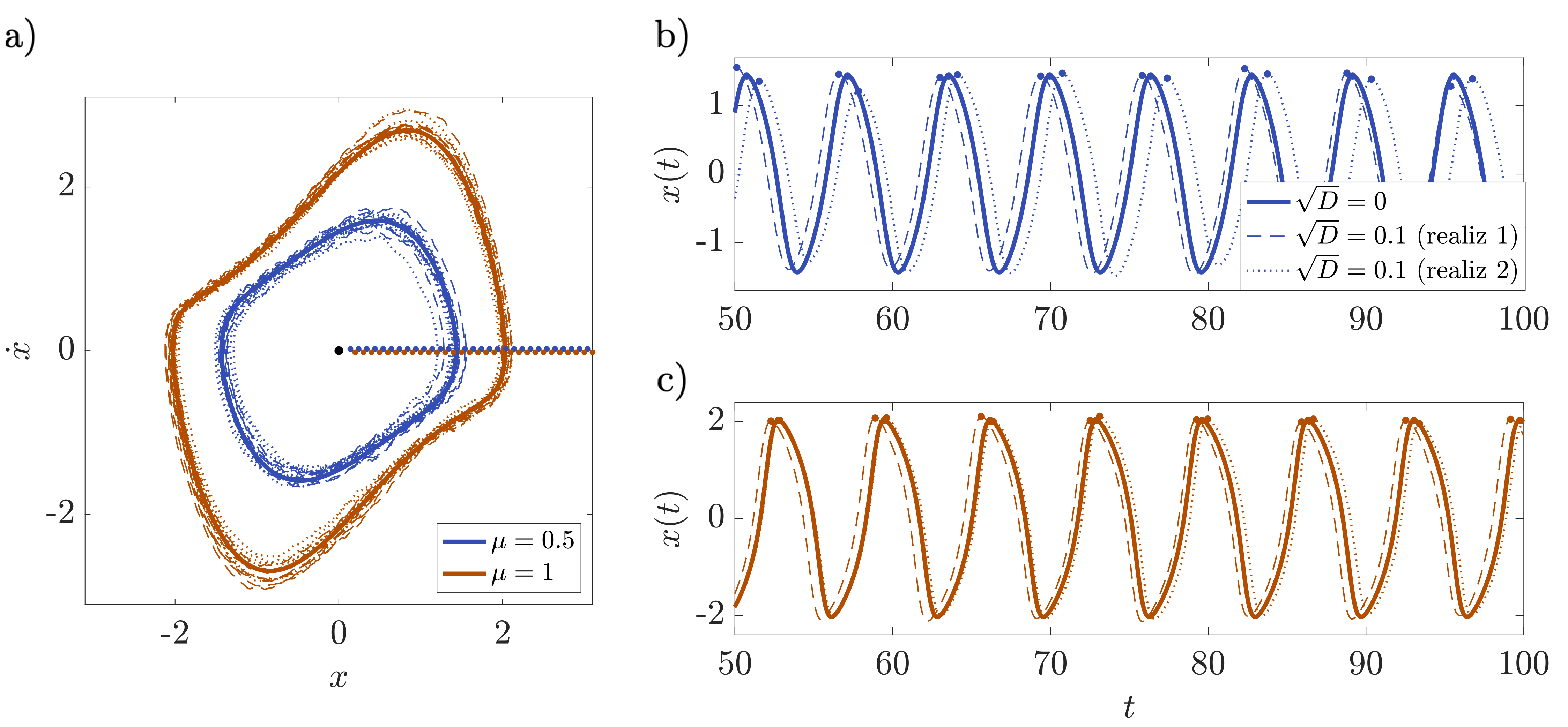}
\caption{{\sf \bf Limit cycle and orbits of vdP oscillator, with and without noise.}
(a) Limit-cycles (solid lines), obtained by numerically integrating a vdP oscillator with $\sqrt{D}=0$, and two noisy orbits (dashed and dotted lines) with  $\sqrt{D}=0.1$, for $\mu = 0.5$ and $\mu = 1$ (see legend). In each case the corresponding Poincar\'e section (with a small vertical displacement for ease of visualization) is highlighted with dots (while an isolated black dot indicates the origin of coordinates). (b) Trajectories $x(t)$ for $\mu = 0.5$ without ($\sqrt{D}=0$) and with noise (two realizations with $\sqrt{D}=0.1$). (c) Trajectories $x(t)$ for $\mu = 1$ without ($\sqrt{D}=0$) and with noise (two realizations with $\sqrt{D}=0.1$) from arbitrary initial conditions. In both (b) and (c) a time interval of 50 units after an initial transient of the same duration is shown, with crossings of the Poincar\'e section highlighted by dots.}
\label{FigS1}
\end{figure}

As the distance to the supercritical Hopf bifurcation given by $\mu$ increases, the van der Pol (vdP) oscillator behaves less and less like a Stuart-Landau (SL) oscillator and gradually more like a relaxation oscillator, going across different regions of its limit cycle (which is conspicuously different from a circle) at different speeds \cite{strogatzbookSM}. See Fig.~\ref{FigS1} (a) for an illustration of the limit cycles (solid lines) corresponding to $\mu = 0.5$ and $1$, the two parameter values under consideration in the main text. We also include two phase-space orbits with noise of the same strength used in our simulations, namely $\sqrt{D}=0.1$ (dashed and dotted lines; note, these noisy orbits are not closed). For further illustration, Fig.~\ref{FigS1} (b) shows numerically-integrated trajectories $x(t)$ over a time interval of $50$ units (after a transient of the same duration from an arbitrary initial condition) for $\mu = 0.5$, again both without noise ($\sqrt{D}=0$) and with noise (the same two realizations with $\sqrt{D}=0.1$), while  panel (c) of the same figure displays analogous results for $\mu = 1$.

Given the behavior just described, the geometric phase given by $\arctan(\dot{x}/x)$ (considering the quadrant) may not be an appropriate dynamical phase variable \cite{pikovskySM}. In the absence of an analytical expression for the isochrones (as far as we know), we define the phase by means of a Poincar\'e section \cite{pikovskySM}. We do not use the phase definition $\arctan(\ddot{x}/\dot{x})$, which has been used in deterministic systems and is based on concepts from the differential geometry of curves \cite{osipovSM}, due to the possibility that the time derivatives may be sensitively affected by noise.

We place the Poincar\'e section at the half line given $\dot{x} = 0$ for $x>\sqrt{\mu}/5$, which is illustrated in Fig.~\ref{FigS1} (a). The two limit cycles intersect that half line approximately at $x \approx 2 \sqrt{\mu}$, so this choice allows for strong perturbations due to noise and coupling to other oscillators. As $\dot{x}$ goes from positive to negative for sufficiently positive $x$, the phase $\phi$ increases by $2\pi$ with respect to the previous crossing of the section. But there is an additional precautionary measure: to avoid detecting spurious return times due to the constant perturbations to which the oscillators in the noisy rings are subjected, we only consider those for which the time elapsed since the previous return time is at least one tenth of the intrinsic period $2\pi$. In this way we avoid taking rapid back-and-forth oscillations across the section for a full period. We have found that our results are robust against moderate variations of these threshold values.

\section{Continuum approximation: phase interface}

Starting from a ring of oscillators with lattice constant $a$ whose state is well captured by the phase profile $\{\phi_j(t)\}_{j=1}^L$, for a relatively slow spatial variation of the phases (as expected for coupling strengths $K$ well above the threshold for synchronization $K_c$), one can perform a continuum approximation like the one discussed in Ref.~\cite{gutierrezSM} for the case of columnar disorder in arbitrary dimensions. As noted in the main text, by neglecting terms of order higher than $\mathcal{O}(a^2)$, which is the dominant order for spatial coupling of the phase field in the ``oscillating medium'', yields the effective continuum equation
\begin{equation}
\partial_t \phi(x,t) = \nu\, \partial_x^2 \phi(x,t) + \frac{\lambda}{2}\, [\partial_x \phi(x,t)]^2 + \text{noise}.
\label{eqkpzSM}
\end{equation}
Here $\nu \equiv a^2\, \Gamma_\phi^{(1)}(0)$ and $\lambda/2 \equiv a^2\, \Gamma_\phi^{(2)}(0)$, where $\Gamma_\phi(\Delta \phi)$ is the coupling function given in terms of the phases, and $\Gamma_\phi^{(k)}(0)$ its $k$-th derivative at the origin. In the modelling of interfacial growth, the term proportional to $\nu$ corresponds to a smoothening surface tension mechanism, while that of $\lambda$ is the Kardar-Parisi-Zhang (KPZ) nonlinearity, representing growth along the local normal direction to the interface \cite{barabasiSM}. We can obtain more detailed expressions for $\nu$ and $\lambda$ for specific oscillator systems:

\begin{itemize}
    \item For the Kuramoto-Sakaguchi (KS) phase-oscillator model, $\Gamma_{\text{KS},\phi}(\Delta \phi) = \Gamma_\text{KS}(\Delta \phi)$. Hence, $\Gamma_\text{KS}^{(1)}(0) = \cos \delta$ while $\Gamma_\text{KS}^{(2)}(0) = -\sin \delta$, so that the relative strength of the KPZ nolinearity with respect to the surface tension is $|\lambda/2\nu| = \tan \delta$.

    \item Concerning SL and vdP oscillators, the phase reduction for very similar parameters as those considered here has been accomplished in Ref.~\cite{gutierrez2SM} in the presence of columnar disorder (i.e.\! randomly asigned intrinsic frequencies of oscillation). Assuming this to be valid for time-dependent noise as well, we find $\Gamma_{\text{SL},\phi}^{(1)}(0) =  1/\sqrt{1+\gamma^2}$ and $\Gamma_{\text{SL},\phi}^{(2)}(0) = - \gamma/\sqrt{1+\gamma^2}$ for SL, and $\Gamma_{\text{vdP},\phi}^{(1)}(0) =  1/2 \sqrt{1+\gamma^2}$ and $\Gamma_{\text{vdP},\phi}^{(2)}(0) = \gamma/2 \sqrt{1+\gamma^2}$ for vdP. In both cases $|\lambda/2\nu| = \gamma$, which provides a meaning to this parameter as the strength of the nonlinearity relative to that of surface tension in the phase-reduced continuum description.
\end{itemize}

The way of relating the microscopic parameters of the oscillators and the coefficients of the continuum equation that we have employed in this section is useful to identify relevant terms in the continuum approximation and, for lack of more detailed information, to motivate the choice of microscopic parameter values that we use in the simulations of the main text. In fact, the latter is based on setting $|\lambda/2\nu| = 5$, which should result in a relatively strong KPZ nonlinearity, and is achieved by choosing $\delta = \atan 5$, in the case of KS, and $\gamma = 5$, for the limit-cycle oscillators. Yet it is quite likely that the full effective Langevin equation of these oscillators in the continuum is much more complex, and only reduces to the KPZ equation asymptotically. If that were the case, the coefficients of the continuum equation would renormalize in a nontrivial way when going towards larger space-time scales, and direct mappings between parameters at different scales such as the ones we here use would yield only rough estimates.

\section{Roughness in lattices of oscillators for different coupling strengths}

Here we briefly explore the behavior of lattices of phase and limit-cycle oscillators as the coupling strength $K$ is varied. Since the focus of this work is on synchronous dynamics, which requires setting the coupling strength above the threshold for synchronization $K_c$, our main objective of this section is to show how the growth of the roughness $W(t)$ [see Eq.\! (2) of the main text] before saturation remains unaffected in its scaling by changes in the coupling strength so long as $K>K_c$. We also explore what happens for nonsynchronous dynamics, $K<K_c$, which, somewhat surprisingly, displays a similarly robust growth behavior, yet with a different exponent (and, obviously, not followed by a saturation plateau).

\begin{figure}[h!]
\includegraphics[scale=0.34]{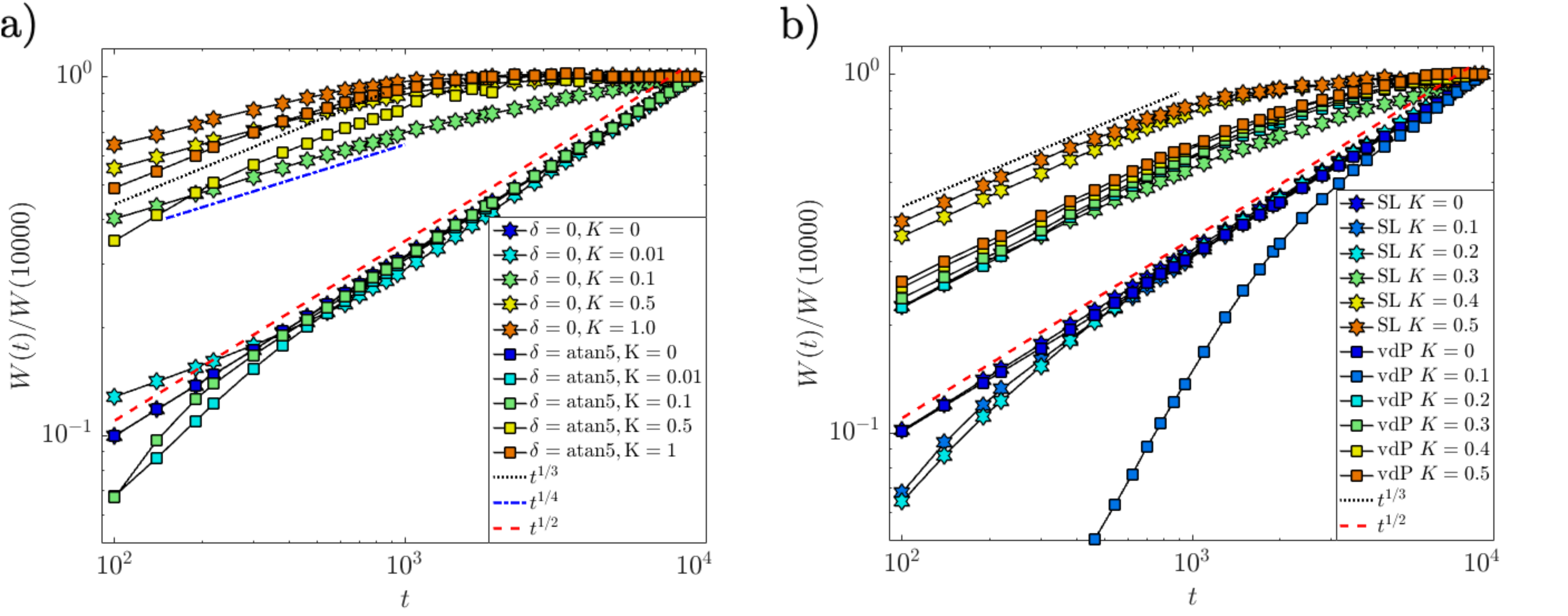}
\caption{{\sf \bf Dependence of the roughness on the coupling strength for lattices of phase and limit-cycle oscillators}
(a) Roughness $W(t)$ for rings of $L=200$ KS oscillators with $\delta = 0$ and $\delta = \atan 5$, for coupling strengths $K= 0, 0.01, 0.1, 0.5$ and $1.0$ (see legend). Power law growths $t^\beta$ with exponents $\beta_{\rm EW} = 1/4,\, \beta_{\rm KPZ} = 1/3$, and $\beta_{\rm RD} = 1/2$ are shown for comparison. Remaining parameters as in the numerical results reported in the main text. (b) Roughness $W(t)$ for rings of $L=200$ SL and vdP oscillators with $\mu = 0.5$, for coupling strengths $K= 0, 0.1 0.2, 0.3, 0.4$, and $0.5$ (see legend). Power law growths $t^\beta$ with exponents $\beta_{\rm KPZ} = 1/3$ and $\beta_{\rm RD} = 1/2$ are shown for comparison. Remaining parameters as in the numerical results reported in the main text. Results in both panels based on $1000$ realizations.}
\label{FigS3}
\end{figure}

In Fig.~\ref{FigS3} (a) we show the roughness of $W(t)$ in systems of $L=200$ KS (phase) oscillators, for the two values of $\delta$ explored in the text, and various values of the coupling strength $K$. Provided that $K>K_c$, the growth prior to saturation is compatible with the $W(t) \sim t^{\beta_{\rm EW}}$ and $\sim t^{\beta_{\rm KPZ}}$ behavior of Edwards-Wilkinson ($\beta_{\rm EW} = 1/4$) and KPZ universality ($\beta_{\rm KPZ} = 1/3$) for $\delta = 0$ and $\atan 5$, respectively, as it was reported in the main text for larger system sizes. While the numerical value of $K_c$ (which must lie between the largest value of $K$ that does not display saturation and the smallest that does) is vastly different depending on $\delta$, the behavior is practically identical for $K<K_c$, at least after an initial transient. Namely, the roughness for such values of $K$ displays a permanent growth of the form $W(t) \sim t^{\beta_{\rm RD}}$, where $\beta_{\rm RD} = 1/2$ is the exponent corresponding to random deposition \cite{barabasiSM}. In fact, for uncoupled KS oscillators, $K=0$, the dynamics consists of a uniform velocity term plus a delta-correlated time-dependent noise, which essentially is a random-deposition process. What is perhaps not so obvious is that no appreciable changes are observed for larger $K$ as long as $0<K<K_c$, for either value of $\delta$.

In Fig.~\ref{FigS3} (b) we also show the roughness of $W(t)$ corresponding to various values of $K$, this time for systems of $L=200$ SL and vdP (limit-cycle) oscillators for $\mu = 0.5$. Note that, while the color code in both Fig.~\ref{FigS3} (a) and (b) is such that dark blue corresponds to $K=0$ and orange to the largest value of $K$ considered, in each case it is adapted to the range of values considered. We find that the growth for $K>K_c$ always follows the KPZ scaling $W(t) \sim t^{\beta_{\rm KPZ}}$, as expected, while for $K<K_c$ the system eventually settles into what appears to be a permanent growth that follows the random-deposition scaling $W(t) \sim t^{\beta_{\rm RD}}$, as in the case of phase oscillators, which again is expected for $K=0$.


\section{Correlations in Fourier space: Structure factors of phase profiles}

A correlation function that we have not considered in the main text, yet is frequently studied in the physics of kinetic roughening \cite{siegert96SM,lopezphysaSM}, is the power spectral density of phase fluctuations,
\begin{equation}
S(k,t) \equiv \langle \hat{\phi}(k, t)\hat{\phi}(-k, t)\rangle = \langle |\hat{\phi}(k, t)|^2\rangle,
\label{Sk}
\end{equation}
where $\hat{\phi}(k,t)$ is the (1D) space Fourier transform of the (discretized) phase field $\phi(x,t)$ and $k$ is the wavenumber. In fact, $S(k,t)$ is analytically related with the real-space correlations $G(\ell,t)$ via space Fourier transforms \cite{krug97SM}. In kinetic-roughening systems for which the Family-Viscsek (FV) dynamic scaling ansatz holds \cite{Family1985SM,barabasiSM,halpinhealySM,krug97SM}, this correlation function (based on local heights above a substrate instead of local dynamical phases) satisfies the scaling form
\begin{equation}
S(k,t) = k^{-(2  \alpha +1)} s(k t^{1/z}),
\label{Skscal}
\end{equation}
where the scaling function $s(y) \sim y^{2\alpha +1}$ for $y\ll 1$, while it becomes constant for $y\gg 1$.

\begin{figure}[h!]
\includegraphics[scale=0.34]{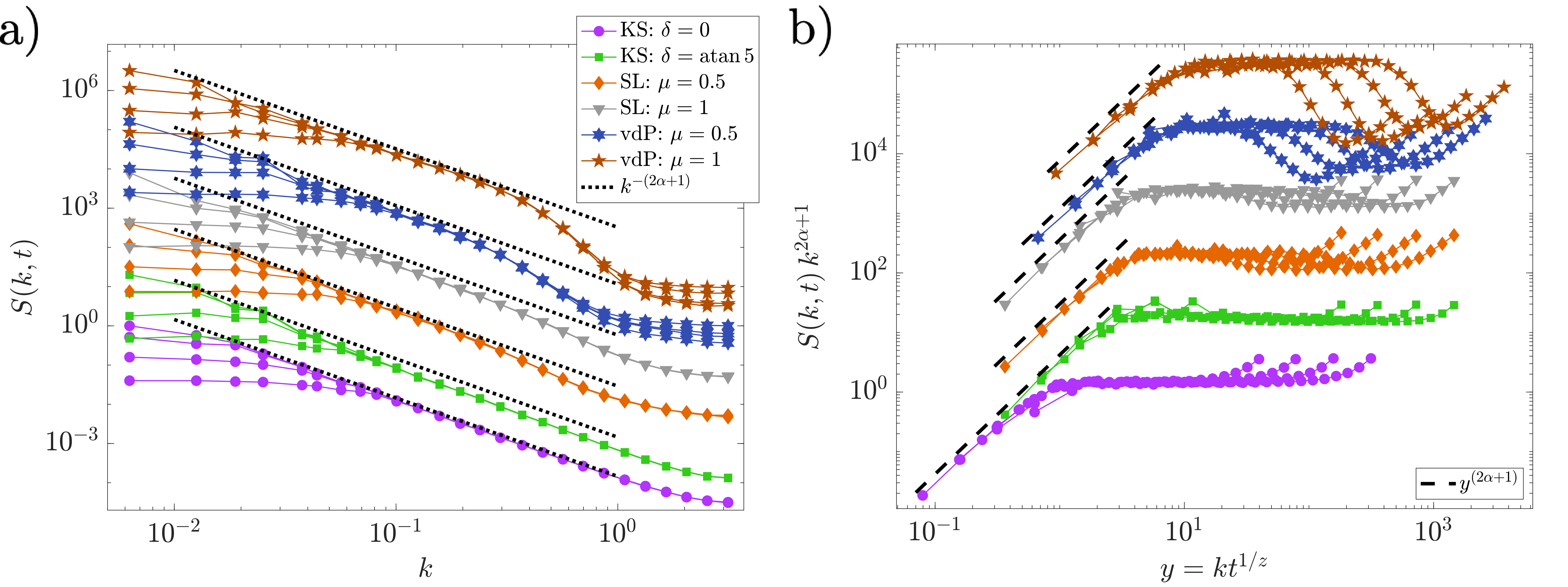}
\caption{{\sf \bf Phase structure factors of rings of self-sustained oscillators.}
(a) Structure factor $S(k,t)$ at four instants of time corresponding to $t/t^* = 1/8^z, 1/4^z, 1/2^z$ and $1$ (data for later times lie above those for earlier times) for rings of KS, SL, and vdP oscillators, with the same parameter choices as in the figures of the main text (see legend). Dotted lines are proportional to $k^{-(2\alpha + 1)}$. (b) Rescaled structure factor based on the FV dynamic scaling Ansatz in Eq.~(\ref{Skscal}). Dashed lines are proportional to $y^{(2\alpha + 1)}$, where $y= k t^{1/z}$. In both panels, the exponent values are $\alpha = \alpha_{\rm KPZ} = \alpha_{\rm EW} = 1/2$, in all cases, and $z = z_{\rm KPZ} = 3/2$, except for the KS ring with $\delta = 0$, for which $z=z_{\rm EW}=2$, and an arbitrary vertical displacement has been performed for visualization purposes. Results based on $5000$ realizations of the noise.}
\label{FigS2}
\end{figure}

In Fig.~\ref{FigS2} (a) we show the structure factor $S(k,t)$, Eq.\ (\ref{Sk}), for the same KS, SL and vdP rings, and parameters studied in the main text, with curves corresponding to later times appearing above those for earlier times, as they saturate at a smaller $k \sim \xi(t)^{-1}$. As in the case of the real-space phase-difference correlation functions $G(\ell,t)$ shown in Fig.~2 of the main text, the four instants of time are chosen so that $t/t^* = 1/8^z, 1/4^z, 1/2^z$ and $1$, which roughly corresponds to $\xi(t)/L \sim 1/8, 1/4, 1/2$ and $1$.

The validity of the FV dynamic scaling ansatz given in Eq.~(\ref{Skscal}) is illustrated in Fig.~\ref{FigS2} (b), which shows an excellent collapse for the rescaled structure factor up to wavenumbers close to the cutoff in all cases, except for the very small space-time scales of the vdP oscillators (as happens for the real-space results reported in Fig.~2 of the main text). Note moreover the excellent agreement with the dashed line in panel (a) displaying the expected power-law dependence $k^{-(2\alpha+1)}$ for wavenumbers $k\gg t^{-1/z}$. In the case of KS rings with $\delta = 0$, the 1D EW exponents, $\alpha = \alpha_{\rm EW} = 1/2$ and $z=z_{\rm EW}=2$, have been employed, while in every other case the values of choice are those of the 1D KPZ universality class, namely, $\alpha =\alpha_{\rm KPZ}= 1/2$ and $z = z_{\rm KPZ}=3/2$, just as in the results reported in the main text.


\begin{thebibliography}{60}
\bibitem [{note2()}]{note2}%
  \BibitemOpen
  \href@noop {} {} \bibinfo {note} {To mitigate EW
  crossover effects \cite{forrest}, we use a relatively strong $|\lambda/2\nu|
  = 5$ by setting $\delta = \atan 5 \approx 1.37$ for the KS model, and $\gamma
  = 5$ for SL and vdP oscillators; see the SM \cite{sm}.}\BibitemShut {Stop}%
\bibitem [{notecoeffs()}]{notecoeffs}%
  \BibitemOpen
  \href@noop {} {}\bibinfo {note} {Note that the relation
  between the microscopic parameters of the oscillators and the coefficients of
  Eq.\ (7), while useful to identify relevant terms in the continuum
  approximation, is unlikely to yield accurate results, as nontrivial
  renormalization of the coefficients is generally expected.}\BibitemShut
  {Stop}%
\end{thebibliography}
\end{document}